# High Temperature Far Infrared Dynamics of Orthorhombic NdMnO$_3$: Emissivity and Reflectivity

Néstor E. Massa*,[1] Leire del Campo,[2] Domingos De Sousa Meneses,[2] Patrick Echegut,[2] María Jesús Martínez-Lope,[3] and José Antonio Alonso[3]

[1] Laboratorio Nacional de Investigación y Servicios en Espectroscopía Óptica-Centro CEQUINOR, Universidad Nacional de La Plata, C.C. 962, 1900 La Plata, Argentina.

[2] CNRS, CEMHTI UPR3079, Univ. Orléans, F-45071 Orléans, France

[3] Instituto de Ciencia de Materiales de Madrid, CSIC, Cantoblanco, E-28049 Madrid, Spain.

•e-mail: neemmassa@gmail.com

# ABSTRACT

We report on near normal far- and mid-infrared emission and reflectivity of NdMnO$_3$ perovskite from room temperature to sample decomposition above 1800 K. At 300 K the number infrared active phonons is in close agreement with the 25 calculated for the orthorhombic $D_{2h}^{16}$-Pbnm (Z=4) space group. Their number gradually decreases as we approach the temperature of orbital disorder at ~1023 K where the orthorhombic O′ lower temperature cooperative phase coexists with the cubic orthorhombic O. At above ~1200 K, the three infrared active phonons coincide with the expected for cubic Pm-3m (Z=1) in the high temperature insulating regime.

Heating samples in dry air triggers double exchange conductivity by Mn$^{3+}$ and Mn$^{4+}$ ions and a small polaron mid-infrared band. Fits to the optical conductivity single out the octahedral antisymmetric and symmetric vibrational modes as main phonons in the electron-phonon interactions at 875 K. For 1745 K, it is enough to consider the symmetric stretching internal mode. An overdamped defect induced Drude component is clearly outlined at the highest temperatures. We conclude that Rare Earth manganites e$_g$ electrons are prone to spin, charge, orbital, and lattice couplings in an intrinsic orbital distorted perovskite lattice favoring embryonic low energy collective excitations.

**PACS**

Collective effects 71.45.-d

Infrared spectra 78.30.-j

Multiferroics 75.85.+t

Emissivity 78.20.Ci (far infrared)

# INTRODUCTION

The family of RMnO$_3$ (R=Rare Earths) has been thoroughly studied during past decades as parents of mixed valence perovskites. They have a myriad of interesting properties ranging from colossal magnetoresistance to electron ordering below a critical temperature. NdMnO$_3$, in particular, has attracted a lot of attention because it generates compounds with rich phase diagrams that may result in doping-enhance lower dimensionalities. In these systems, electric dipoles and magnetic moments may couple fulfilling the primary requirement for magnetoelectrics in which ferroelectricity and magnetism coexist in the so-called multiferroic phases.[1]

NdMnO$_3$ is an insulator due to the absence of mixed valences states. [2] It belongs to the group of compounds with rare earth with largest ionic radius for which colossal magnetoresistance has been reported. [3] Nd occupies an intermediate place between La and Eu potentially increasing the topological metastability of perovskite lattice as it approaches the hexagonal non-perovskite arrangement found by introducing smaller ions.[2] Its inclusion decreases the Mn-O-Mn angle yielding a room temperature O'-orthorhombic phase belonging to the space group $D_{2h}^{16}$-P*bnm*..[4] The low temperature antiferromagnetic phase of NdMnO$_3$ at $T_N$~78 K is characterized by ferromagnetic alignment of the Mn moments in the ***ab*** plane. The 2-fold degenerated e$_g$ orbital breaks down and stabilizes the A–type antiferromagnetic Mn ordering.[5]

At higher temperatures, between ~1023 K and ~1273 K, two substructures, the so called O-orthorhombic (octahedral cooperative buckling) and the O'-orthorhombic (octahedral cooperative buckling and Jahn-Teller (JT) distortions) phases coexist within a disordered orbital two-phase region. [6] Above ~1273K, the Jahn-Teller (JT) ordering distortion of $MnO_6$ octahedra goes undetected as in net metrically cubic lattice underlying a possible dynamical JT effect.[7,8]

Below ~973 K, the JT distortion prevails in an ordered phase that alternates staggered pattern of $d_{3x2-r2}$ and $d_{3y2-r2}$ orbitals in the **_ab_** plane repeating itself along the *c* axis. The allowed $Q_2$ and $Q_3$ JT distortions compete with octahedral rotations cooperatively with marked sublattice basal plane deformations of the $GdFeO_3$-type structure. [9]

Here, we report on near normal far- and mid-infrared emission and reflectivity measurements of $NdMnO_3$ from 300 K to sample decomposition. After reviewing the experimental background we introduce our high temperature data putting emphasis in the orbital disordered phase that it is fundamental for understanding $NdMnO_3$ magnetoelectric couplings and phonon anharmonicities at low temperatures.[10]

We have also found that heating $NdMnO_3$ in dry air activates a mid-infrared band and double exchange conductivity by $Mn^{3+}$ and $Mn^{4+}$ ions as for in divalent $A^{2+}$ substituted manganites.[11] This yields a scenario by which small polaron fits to the optical conductivity single out the octahedral antisymmetric and symmetric vibrational modes as main phonons in the electron-phonon interactions at 875 K. For 1745 K, it is enough to consider the symmetric stretching internal mode. We also detect above ~1200 K an emerging overdamped defect induced Drude response characteristic of poor conducting oxides that is clearly delineated coexisting with the polaronic broad mid-infrared band. [11] At these high temperatures the number of infrared active modes coincides with the predicted for the cubic space group Pm-3m (Z=1)(Ref. 12).

## EXPERIMENTAL DETAILS

Polished high quality samples in the shape of 10mm diameter $NdMnO_3$ pellets were prepared from polycrystalline powders obtained by soft chemistry procedures.

Stoichiometric amounts of $Nd_2O_3$ and $MnCO_3$ were dissolved in citric acid where droplets of concentrated $HNO_3$ were added to favor the solution of the oxide. The citrate + nitrate solution was slowly evaporated leading to an organic resin containing a random distribution of cations. It was first dried at 120 °C and then slowly decomposed up to 600 °C. All organic material and nitrate were eliminated in a subsequent treatment during 12 hrs at 700 °C in air. This yielded a highly reactive precursor that was then treated at 1100 °C in a $N_2$ flow during 12 hrs. The sample was annealed in an inert atmosphere in order to avoid the formation of oxidized $NdMnO_{3+\delta}$, containing a significant amount of $Mn^{4+}$. $NdMnO_3$ was thus obtained as a well-crystallized powder. The crystal structure at ambient temperature, refined from X-ray diffraction data (Fig. 1), corresponds to the conventional orthorhombic space group $D_{2h}^{16}$-$P_{bnm}$. The refined unit cell parameters are a=5.4294(8), b=5.8438(8), c=7.574(1) Å in good agreement with published data [2].

Temperature dependent medium (MIR), and far infrared (FIR) near normal reflectivity from 300 K to 850 K was measured with a FT-IR Bruker 113v interferometer at 2 cm$^{-1}$ resolution using a reflectance accessory. A liquid He cooled bolometer and a deuterated triglycine sulfate (DTGS) detector were employed to completely cover the spectral range of interest. A gold mirror was used as 100% reflectivity reference since, as we show below (Fig. 3),1-emissivity and reflectivity spectra are in excellent agreement. This makes possible the use of the same sample with both techniques without altering its surface (by running first reflectivity in vacuum and then emissivity in dry air). To avoid the effects of interference fringes due to the semi-transparency of NdMnO3 in the THz region we only used 2 mm or thicker samples being the acceptance criteria their no detection on the band profiles.

Sample and reference were placed in the sample chamber inside the spectrometer in equivalent positions as related to the placement of the detector and the ratio between sample and reference signals gave us the reflectivity spectra For high temperature reflectivity (up to ~850 K) we used a heating plate adapted to the near normal reflectivity attachment in the Bruker 113v vacuum chamber. In this temperature range, the spurious infrared signal introduced by the hot sample thermal radiation was corrected to obtain the reflectivity values. All our measurements were taken on heating runs.

Phonon normal emissivity was measured with two Fourier transform infrared spectrometers, Bruker Vertex 80v and Bruker Vertex 70, which are optically coupled to a rotating table placed inside a dry air box (see ref. 13 for device description and scheme). The actual experimental set up is a modification of the device thoroughly described in [14]. The use of two spectrometers allows us to simultaneously measure the spectral emittance in two dissimilar spectral ranges from 40 cm$^{-1}$ to 7000 cm$^{-1}$. The sample, which was heated with a 500 W pulse Coherent $CO_2$ laser, was positioned on the rotating table at the focal point of both spectrometers in a position equivalent to that of internal radiation sources inside the spectrometers. In this measuring configuration, the sample, placed outside the spectrometer, is the infrared radiation source, and conversely, the sample chamber inside the spectrometers is empty.

We will now briefly review the background of emission measurementsThe normal spectral emissivity of the sample, $E(\omega,T)$, is given by the ratio of its luminescence ($\mathscr{L}_S$) relative to the black body's luminescence ($\mathscr{L}_{BB}$) at the same temperature T and geometrical conditions, thus,

$$E(\omega,T) = \frac{L_S(\omega,T)}{L_{BB}(\omega,T)} \qquad (1)$$

In practice, the evaluation of this quantity needs the use of a more complex expression because the measured fluxes are polluted by parasitic radiation. This is because part of the spectrometer and detectors are at 300 K. To eliminate this environmental contribution the sample emissivity is retrieved from three measured interferograms,i.e., sample, $I_S$ ; black body, $I_{BB}$ ; and, environment, $I_{RT}$ ; by applying the following relation [15],

$$\boldsymbol{E}(\omega,T) = \frac{FT(I_S - I_{RT})}{FT(I_{BB} - I_{RT})} \times \frac{\boldsymbol{P}(T_{BB}) - \boldsymbol{P}(T_{RT})}{\boldsymbol{P}(T_S) - \boldsymbol{P}(T_{RT})} \boldsymbol{E}_{BB} \qquad (2)$$

where *FT* stands for Fourier Transform., and I for measured interferograms i.e., sample, $I_S$ ; black body, $I_{BB}$ ; and, environment, $I_{RT}$ . $\mathscr{P}$ is the Planck's function taken at different temperatures T; i.e., sample, $T_S$; blackbody, $T_{BB}$; and surroundings, $T_{RT}$. $\boldsymbol{E}_{BB}$ is a correction that corresponds to the normal spectral emissivity of the black body reference (a LaCrO$_3$ Pyrox PY 8 commercial oven) and takes into account its non-ideality. [15]

Emissivity allows contact free measurement of temperature of an insulator using the Christiansen frequency; i.e., the frequency where the refraction index is equal to one, and the extinction coefficient is negligible, just after the highest longitudinal optical phonon frequency. The temperature is calculated using eq.

(2) with the emissivity $E(\omega,T)$ at that frequency set equal to one. At higher temperatures, we have also used a thermocouple since the Christiansen inflection is absent in conducting samples.

After acquiring the optical data we placed all our spectra in a more familiar near normal reflectivity framework by noting that

$$R = 1 - E \qquad (3)$$

where $R$ is the sample reflectivity. This allows computing phonon frequencies using a standard multioscillator dielectric simulation. [16] We use a description of the dielectric function, $\varepsilon(\omega)$, given by

$$\varepsilon(\omega) = \varepsilon_1(\omega) - i\varepsilon_2(\omega) = \varepsilon_\infty \prod_j \frac{(\omega_{jLO}^2 - \omega^2 + i\gamma_{jLO}\omega)}{(\omega_{jTO}^2 - \omega^2 + i\gamma_{jTO}\omega)} \qquad (4)$$

$\varepsilon_\infty$ is the high frequency dielectric constant taking into account electronic contributions; $\omega_{jTO}$ and $\omega_{jLO}$, are the transverse and longitudinal optical mode frequencies and $\gamma_{jTO}$ and $\gamma_{jLO}$ their respective damping. When needed, we also added the Drude term (plasma contribution) to the dielectric simulation as

$$-\frac{\left(\omega_{pl}^2 + i \cdot (\gamma_{pl} - \gamma_0) \cdot \omega\right)}{(\omega - i\gamma_0) \cdot \omega} \qquad (5)$$

where $\omega_{pl}$ is the plasma frequency, $\gamma_{pl}$ its damping, and $\gamma_0$ is understood as a phenomenological damping introduced to reflect lattice drag effects. When these two dampings are set equal, one retrieves the classical Drude formula.[17]

The real ($\varepsilon_1(\omega)$) and imaginary ($\varepsilon_2(\omega)$) part of the dielectric function (complex permitivity, $\varepsilon^*(\omega)$) is then estimated from fitting [18] the data using the reflectivity R given by

$$R(\omega) = \left| \frac{\sqrt{\varepsilon^*(\omega)} - 1}{\sqrt{\varepsilon^*(\omega)} + 1} \right|^2 \qquad (6)$$

We also calculated the real part of the temperature dependent optical conductivity, $\sigma_1(\omega)$,[19] given by

$$\sigma_1(\omega) = \frac{\omega \cdot \varepsilon_2}{4\pi}. \qquad (7)$$

## RESULTS AND DISCUSSION

Fig. 2 shows the $NdMnO_3$ phonon reflectivity spectrum at room temperature. Its inset shows a smooth band originating in collective excitations active in the THz region. It locks into two concomitant soft modes at $T_N$ that grow in intensity and definition as the temperature is lowered down to 4 K. [10] Here, and although our reflectivities suggest that the 300 K smooth band persists at higher temperatures, we will not proceed this point further because of our high temperature instrumental limitation at very low frequencies.

The evolution of the near normal far infrared emissivity and reflectivity phonon spectra of $NdMnO_3$ up to 1200 K is shown in Fig. 3. On increasing temperature, the set of reflectivity measurements obtained with e interferometer vacuum chamber are compared against taken with the same sample heated inside the emission dr air chamber. Fig. 3 displays the excellent absolute correspondence between the phonon spectra taken under near normal reflectivity and those by emission in the lower temperature range.

The reflectivity of the vacuum heated sample (Fig. 3a) shows phonon peak position softening and band broadening due to lattice anharmonicities. Phonon bands measured by emission and plotted as 1-emissivity (Fig. 3b) have the same frequency softening and band broadening up to about 800 K. Relative weaker bands, at ~250 cm$^{-1}$ and ~450 cm$^{-1}$, show smooth broadening merging into the three bands found increasing the temperature. This effect suggests that the change into an orbital disordered phase expected between 1023 K (where we find band count already as for the three expected in the cubic phase) and 1273 K [6] takes place gradually instead of at a well-defined temperature range. The overall heating temperature dependent effects are fully reversible.

A multioscillator fit (eq. 4) to the phonon reflectivity spectrum at 300 K (Fig. 2), results in 24 active modes (Table I) out of the 25 predicted for the orthorhombic space group Pbnm ($D^{16}_{2h}$-Z=4 ) [4]

$$\Gamma_{IR}(O') = 9B_{1u} + 7B_{2u} + 9B_{3u} \qquad (8)$$

At 873 K only 17 oscillators (table I) are needed for an excellent fit suggesting a higher temperature likely dynamic net increase in lattice symmetry. I.e., although at high temperatures octahedra distortion goes undetected, large thermal factors might be found for oxygens as in $LaMnO_3$ indicating the presence of an underlying dynamical JT effect [2,7]

In addition, starting at about 500 K, the spectra obtained from samples heated in dry air (Fig. 4a) undergo changes at mid-infrared frequencies.

At 668 K a broad band centered at 2000 cm$^{-1}$ emerges from the background. Its profile becomes less defined at ~900 K due to carriers increased mobility, and at higher temperatures, it turns to be almost undefined due to hopping conductivity by $e_g$ electrons triggering an incipient insulator to metal transition due to small polaronic charge carriers. An overdamped zero frequency centered Drude-Lorentzian and its corresponding tail (Fig. 4a, Table I) now detected as in doped conducting oxides.

Known from annealing oxides in air, the freer conducting electrons are consequence of the oxidation process $Mn^{3+} \rightarrow Mn^{4+}$ + 1e- which is equivalent to divalent $A^{2+}$ substitution.[2] Here, the occurrence of mixed $Mn^{3+}$-$Mn^{4+}$ valence gives rise to the double exchange mechanism [20] in a higher-mobility-high-temperature small polaron-like scenario. [11] In our case, since the insertion of extra-oxygen in a perovskite lattice is not possible, nonstoichometry is generated by Nd and Mn vacancies. [2]

Phonon screening by delocalized eg electrons and stronger vibrational damping prevail in an orbital perturbed environment of coexisting O and O´ phases, The interval where orbital disorder takes place is broaden by ~200 K due to the oxidation mechanism.[8]

Above ~1200 K the O-orthorhombic cubic phase is detected [2,8] for which only three oscillators (Table I) are needed to fit our highest temperature emissivity spectra (Fig. 4). They are assigned to the three infrared active phonons of the reduced representation Γ= 3$F_{1u}$ + 1 $F_{2u}$ predicted for the high temperature cubic space group P*m-3m* (Z=1).(pseudocubic P*mcm* Z=1) [12]
Overall, we find that our 1-emissivity spectra show remarkable similarities with the known behavior bulk disordered conducting oxides undergoing a metal to insulating phase transition (Fig. 7 in [21]). It suggests that a quantitative analysis of the high temperature mid-infrared real part of the optical conductivity, eq.(7), within a small polaron context, may help to identify phonon groups involved in main carrier-phonon interactions. The coexistence of localized and itinerant carriers promoted by the strong interactions between charge carriers and ions yields the formation of polarons in highly polarizable distorted oxygen orbitals.

For this purpose, we use the theoretical formulation for small polarons due to nondiagonal phonon transitions as proposed by Reik and Heese.[22,23] In this model, optical properties are to carriers in one small band and interband transitions are excluded. Starting with a Holstein´s Hamiltonian, [24] the frequency dependent conductivity is calculated using Kubo´s formula.[25] Then, the real part of the optical conductivity for finite temperature, $\sigma_1(\omega,\beta)$ is given by

$$\sigma_1(\omega,\beta) = \sigma_{DC} \frac{\sinh\left(\frac{1}{2}\hbar\omega\beta\right)\exp\left[-\omega^2\psi^2 r(\omega)\right]}{\frac{1}{2}\hbar\omega\beta\left[1+(\omega\psi\Delta)^2\right]^{1/4}}, \quad (9)$$

$$r(\omega) = \left(\frac{2}{\omega\Psi\Delta}\right)\ln\left\{\omega\Psi\Delta + \left[1+(\omega\Psi\Delta)^2\right]^{1/2}\right\} - \left[\frac{2}{(\omega\Psi\Delta)^2}\right]\left\{\left[1+(\omega\Psi\Delta)^2\right]^{1/2} - 1\right\}, \quad (10)$$

with $\Delta = 2\varpi_j \Psi$ \hfill (11)

and $\Psi^2 = \dfrac{\left[\sinh\left(\frac{1}{2}\hbar\varpi_j\cdot\beta\right)\right]}{2\varpi_j^2 \eta}.$ \hfill (12)

The conductivity, $\sigma_1(\omega,\beta)$, $\beta=1/kT$, is mainly three parameter dependent; $\sigma_{DC}=\sigma(0,\beta)$, the electrical DC conductivity; the frequency $\varpi_j$ that corresponds to the average between the transverse and the longitudinal optical mode of the $j^{th}$ restrahlen band; and $\eta$, a parameter characterizing the strength of the electron-phonon interaction, i.e., the average number of phonons that contribute to the polarization around a localized polaron. Among all these parameters, $\eta$, is the only free parameter in the optical conductivity computation for each phonon frequency $\varpi_j$. This is because phonon frequencies are fixed by the reflectivity (or 1-emissivity) measurements and the DC-zero frequency-conductivity is known from independent, either optical or transport measurements. [26] $\eta$ ~3 implies a low to mild electron-phonon interaction while a value around 14 would be corresponding to the very strong end. [27]

We assume that frequency dependent conductivities result from the addition of gaussian-like (eq. 9) individual contributions, each of them, calculated at a phonon frequency $\varpi_j$ and a temperature T. The fits are shown in Fig. 5 (table II). The conductivities were also constrained to positive values. They are in agreement with the experimental data. A good fit at 873 K requires only allowing the first-order internal vibrational modes of $NdMnO_3$. In spite of omitting an explicit correlation between the different contributions, the fitted frequencies are very close to the experimental ones, and $NdMnO_3$ an insulator, at 873 K the electron-phonon parameters $\eta_j$ (j=1,2,3), (table II), are in a regime for stronger localization. These values for $\eta$'s, and the individualization of only fundamental lattice frequencies with strong. [28]

In contrast, when temperature is raised up to 1745 K, in the cubic phase, the highest frequency mode at ~570 cm$^{-1}$, and its overtone at ~1140 cm$^{-1}$, are singular to the fit at mid-infrared frequencies. The fundamental vibration band corresponds to the octahedral breathing symmetric stretching mode that may have the total polarization enhanced by orbital disorder.[29] The electron-phonon parameters $\eta_j$ (j=1,2) (Table II), are reduced by the increased electron mobility. The detected overtone implies stronger dipole moments generated by polaronic coupling between the charge and the vibrating lattice. [11]

The behavior at high temperature is also coincident with earlier conclusions from thermal expansion measurements of non-stoichiometric $NdMnO_{3+\delta}$ showing large irreversible lattice anomalies on heating and cooling in the ~1100 K temperature range.[30]

We conclude from our high temperature measurements of $NdMnO_3$ that one may expect, for Rare Earth manganites, eg electrons prone to strong spin-phonon couplings in an intrinsic orbital distorted perovskite lattice favoring embryonic low energy collective excitations at all temperatures.

# CONCLUSIONS

Summarizing, we presented temperature dependent near normal far-infrared emissivity and reflectivity spectra of $NdMnO_3$ from 300 K to 1800 K. We found that when heated in air, as it is known for divalent doping in manganites, oxygen excess triggers hopping conductivity within the double exchange mechanism framework. Close agreeing with the expected number of 25 room temperature phonons for the space group P*bnm* ($D^{16}_{2h}$-Z=4), the number of bands are reduced gradually up to 1200 K where only remain three infrared active phonons associated with the allowed ones for the cubic high temperature insulating phase space group P*m-3m* (Z=1).

Overall, our measurements give a comprehensive view of the evolution of eg electrons at high temperatures in $NdMnO_3$ entangled d-orbitals. The environment put in evidence in emissivity measurements allows inferring the origin of electronic induced mechanisms for colossal magnetoresistance or polar ordering in transition metal oxides involving orbital/charge and/or spin fluctuations. [30-32]

# ACKNOWLEDGEMENTS

N.E.M. is grateful to the CNRS-C.E.M.H.T.I. laboratory and staff in Orléans, France, for research and financial support in performing far infrared measurements. N.E.M. also acknowledges partial financial support (PIP 0010) from the Argentinean Research Council (Consejo Nacional de Investigaciones Científicas y Técnicas-CONICET). Funding through Spain Ministry of Science and Innovation (Ministerio de Ciencia e Innovación) under Project MAT2010 Nº -16404 is acknowledged by J. A. A. and M. J. M-L.

# TABLE I

Dielectric simulation fitting parameters for NdMnO$_3$

| T (K) | $\varepsilon_\infty$ | $\omega_{TO}$ (cm$^{-1}$) | $\Gamma_{TO}$ (cm$^{-1}$) | $\omega_{LO}$ (cm$^{-1}$) | $\Gamma_{LO}$ (cm$^{-1}$) |
|---|---|---|---|---|---|
| 300 | 2.49 | 71.3 | 32.1 | 74.5 | 13.3 |
| | | 76.6 | 15.9 | 78.3 | 41.3 |
| | | 115.6 | 13.0 | 117.0 | 13.0 |
| | | 140.0 | 38.0 | 142.5 | 40.4 |
| | | 174.3 | 8.2 | 176.8 | 6.5 |
| | | 187.4 | 12.6 | 189.4 | 121.3 |
| | | 188.3 | 119.6 | 199.2 | 7.5 |
| | | 202.5 | 11.1 | 205.5 | 10.6 |
| | | 231.8 | 28.8 | 236.1 | 45.4 |
| | | 256.7 | 75.8 | 274.2 | 20.5 |
| | | 276.4 | 14.4 | 277.7 | 15.9 |
| | | 288.5 | 27.4 | 290.9 | 8.8 |
| | | 291.3 | 12.3 | 314.7 | 21.4 |
| | | 312.7 | 77.8 | 315.9 | 24.2 |
| | | 332.6 | 24.2 | 338.6 | 300.4 |
| | | 339.8 | 38.3 | 142.5 | 40.4 |
| | | 385.6 | 23.8 | 387.7 | 29.6 |
| | | 396.8 | 49.8 | 406.3 | 26.6 |
| | | 441.4 | 54.3 | 445.5 | 30.8 |
| | | 462.3 | 22.1 | 465.6 | 12.8 |
| | | 522.3 | 30.4 | 529.0 | 20.3 |
| | | 556.8 | 30.0 | 559.0 | 53.0 |
| | | 566.6 | 60.07 | 607.0 | 79.9 |
| | | 669.2 | 134.0 | 673.5 | 49.7 |

| T (K) | $\varepsilon_\infty$ | $\omega_{TO}$ (cm$^{-1}$) | $\Gamma_{TO}$ (cm$^{-1}$) | $\omega_{LO}$ (cm$^{-1}$) | $\Gamma_{LO}$ (cm$^{-1}$) |
|---|---|---|---|---|---|
| | | 49.5 | 424.4 | 58.9 | 173.3 |
| | | 89.7 | 86.4 | 102.1 | 103.0 |
| | | 167.0 | 64.4 | 177.2 | 68.0 |
| | | 179.6 | 22.4 | 193.9 | 19.0 |
| | | 236.2 | 55.3 | 256.3 | 52.1 |
| | | 280.3 | 455.0 | 299.3 | 32.3 |
| | | 283.5 | 84.0 | 309.5 | 22.7 |
| | | 309.1 | 22.5 | 323.5 | 57.5 |
| | | 315.7 | 56.8 | 387.1 | 39.2 |
| | | 318.7 | 56.5 | 346.2 | 442.7 |

| T (K) | $\varepsilon_\infty$ | $\omega_{TO}$ (cm$^{-1}$) | $\Gamma_{TO}$ (cm$^{-1}$) | $\omega_{LO}$ (cm$^{-1}$) | $\Gamma_{LO}$ (cm$^{-1}$) |
|---|---|---|---|---|---|
| 873 | 1.07 | 375.6 | 47.6 | 384.7 | 546.5 |
|  |  | 420.8 | 147.6 | 442.6 | 205.7 |
|  |  | 449.8 | 103.8 | 460.8 | 73.7 |
|  |  | 479.5 | 586.3 | 511.6 | 105.8 |
|  |  | 509.8 | 57.8 | 529.2 | 66.1 |
|  |  | 543.9 | 40.1 | 619.6 | 701.1 |
|  |  | 616.6 | 197.0 | 639.1 | 95.8 |
|  |  | 1682.9 | 1702.6 | 1904.4 | 6202.0 |

| T (K) | $\varepsilon_\infty$ | $\omega_{TO}$ (cm$^{-1}$) | $\Gamma_{TO}$ (cm$^{-1}$) | $\omega_{LO}$ (cm$^{-1}$) | $\Gamma_{LO}$ (cm$^{-1}$) |
|---|---|---|---|---|---|
| 1745 | 1.03 | 155.1 | 61.4 | 183.1 | 68.1 |
|  |  | 334.3 | 157.9 | 455.3 | 184.5 |
|  |  | 538.1 | 144.3 | 600.5 | 216.0 |
|  |  | 3603.8 | 6354.4 | 6790.7 | 5267.4 |
|  |  |  |  | $\omega_{pl}$ (cm$^{-1}$) |  |
|  |  |  | 9732.1 | 3658.8 | 10562.7 |

## Table II

Parameters of the small polaron theory fits to the optical conductivity of NdMnO$_3$ at 873 K and 1745 K as described in the text. Note that vibrational frequencies are in agreement with experimental bands (in brackets) and that the conductivity for the highest temperatures may be described by simply considering the electron-phonon interaction associated to the octahedral breathing mode and its overtone. The DC conductivities, $\sigma_{DC}$, were kept fixed.
.

| T K | $\sigma_{DC}$ (ohm$^{-1}$-cm$^{-1}$) | $\eta_1$ | $\varpi_{ph1}$ (cm$^{-1}$) | $\eta_2$ | $\varpi_{ph2}$ (cm$^{-1}$) | $\eta_3$ | $\varpi_{ph3}$ (cm$^{-1}$) |
|---|---|---|---|---|---|---|---|
| 873 | 20.1 | 12.8 | 386.1 (375.6) | 9.1 | 473. (480.0)* | 16.1 | 615.7 (616.6) |
| 1745 | 65.0 | | | 8.2 | 590. (570.0) | 5.0 | 1160. |

(*) Average of poorly defined phonon bands found in this frequency range.

# FIGURE CAPTIONS

**Figure 1** (color online) X-ray (CuKα) diffraction pattern for NdMnO$_3$. The inset corresponds to a view of the orthorhombic perovskite structure.

**Figure 2** (color online) NdMnO$_3$ phonon near normal reflectivity at 300 K; experimental: dots, full line: fit. Inset: Semilog plot of the complete far infrared near normal reflectivity of NdMnO$_3$ at 300 K.

**Figure 3** (color online) Same sample NdMnO$_3$ high temperature (a) near normal reflectivity and (b) 1-emissivity in the vibrational region. Arrows point to vibrational groups denoting the gradual increase of orbital disorder that renders the "O" phase net higher temperature-higher lattice symmetry above ~1100 K. For better viewing the spectra have been displaced vertically by 0.10 relative to each other.

**Figure 4** (color online) (a) Temperature dependent near normal 1-emissivity of NdMnO$_3$ heated in air showing the small polaron band centered at 2000 cm$^{-1}$ and its evolution into an overdamped Drude tail at ~1600 K. The mid-infrared spectra below 700 K have been cut because the sample is only partially absorbent. There are not enough heat-generated defects within the gap and transparency induces high levels of noise. The reflectivity of NdMnO$_3$ heated in vacuum is shown in dashed lines at 300 K and 850 K. For better viewing the spectra have been vertically shifted relative to each other by 0.10; (b) NdMnO$_3$ near normal 1-emissivity at 1745 K. The individual Drude (dots) and mid-infrared band (squares) contributions to the 1-emissivity are outlined; (c) NdMnO$_3$ near normal 1-emissivity at 873 K. The individual mid-infrared band contribution is outlined (squares).

**Figure 5** (color online) Temperature dependent optical conductivity of NdMnO$_3$ at 1745 K and 873 K. Full line: experimental; superposing dots: fit**.** The fit assumes that conductivity is the sum of gaussian-like (eq. 9) contributions (drawn in full lines), each calculated at a phonon frequency $\varpi_j$ and at a temperature T (see text and table II).

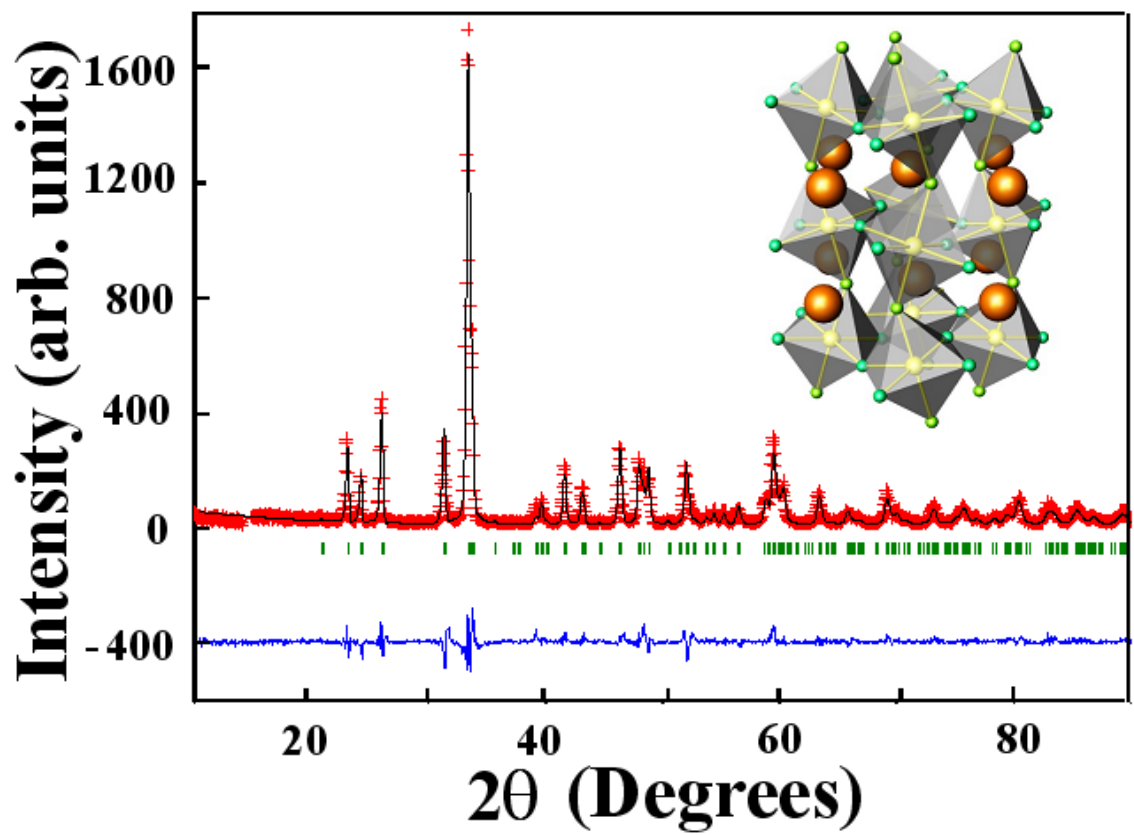

Figure 1
Massa et al

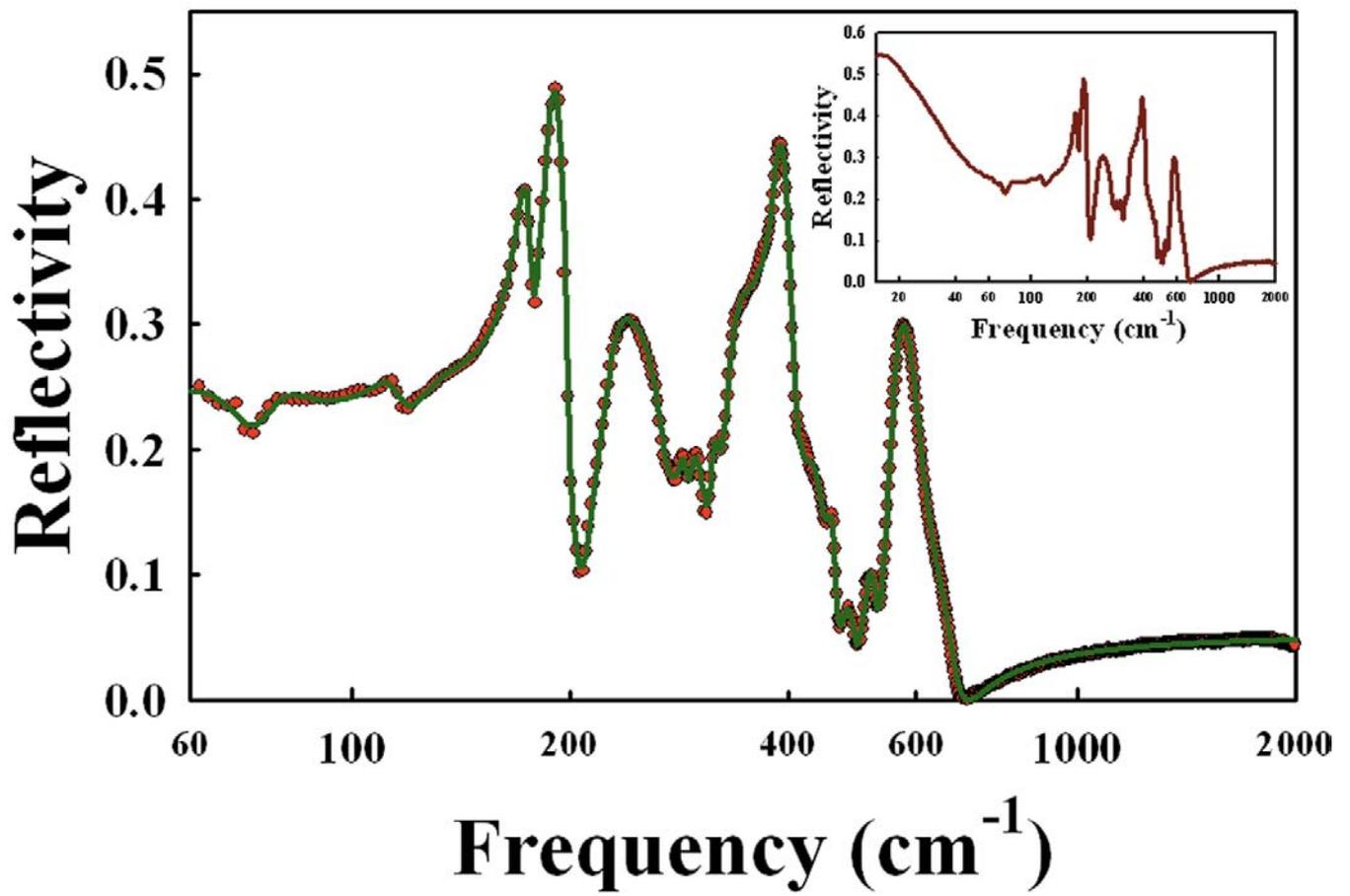

**Figure 2**
**Massa et al**

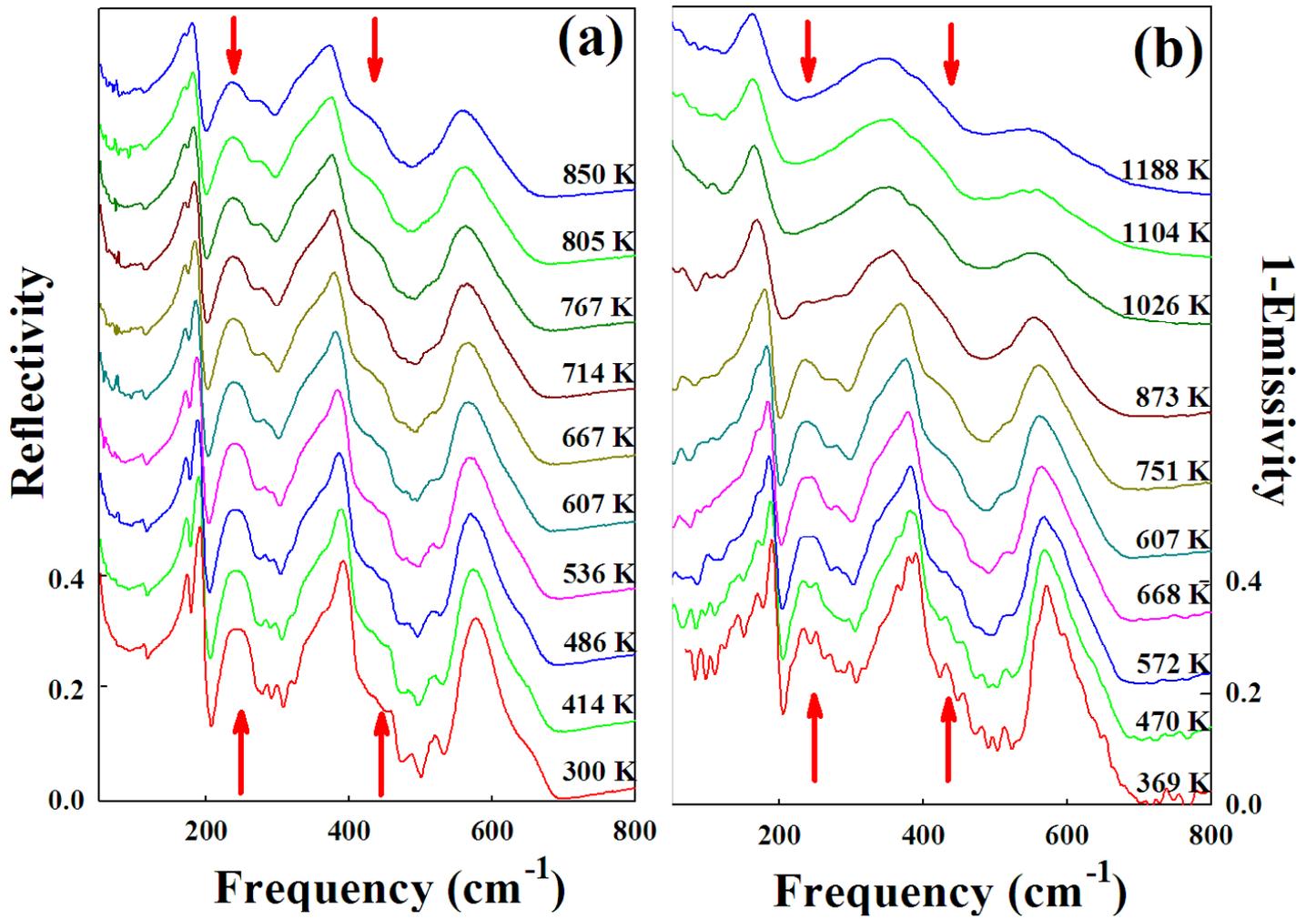

Figure 3
Massa et al

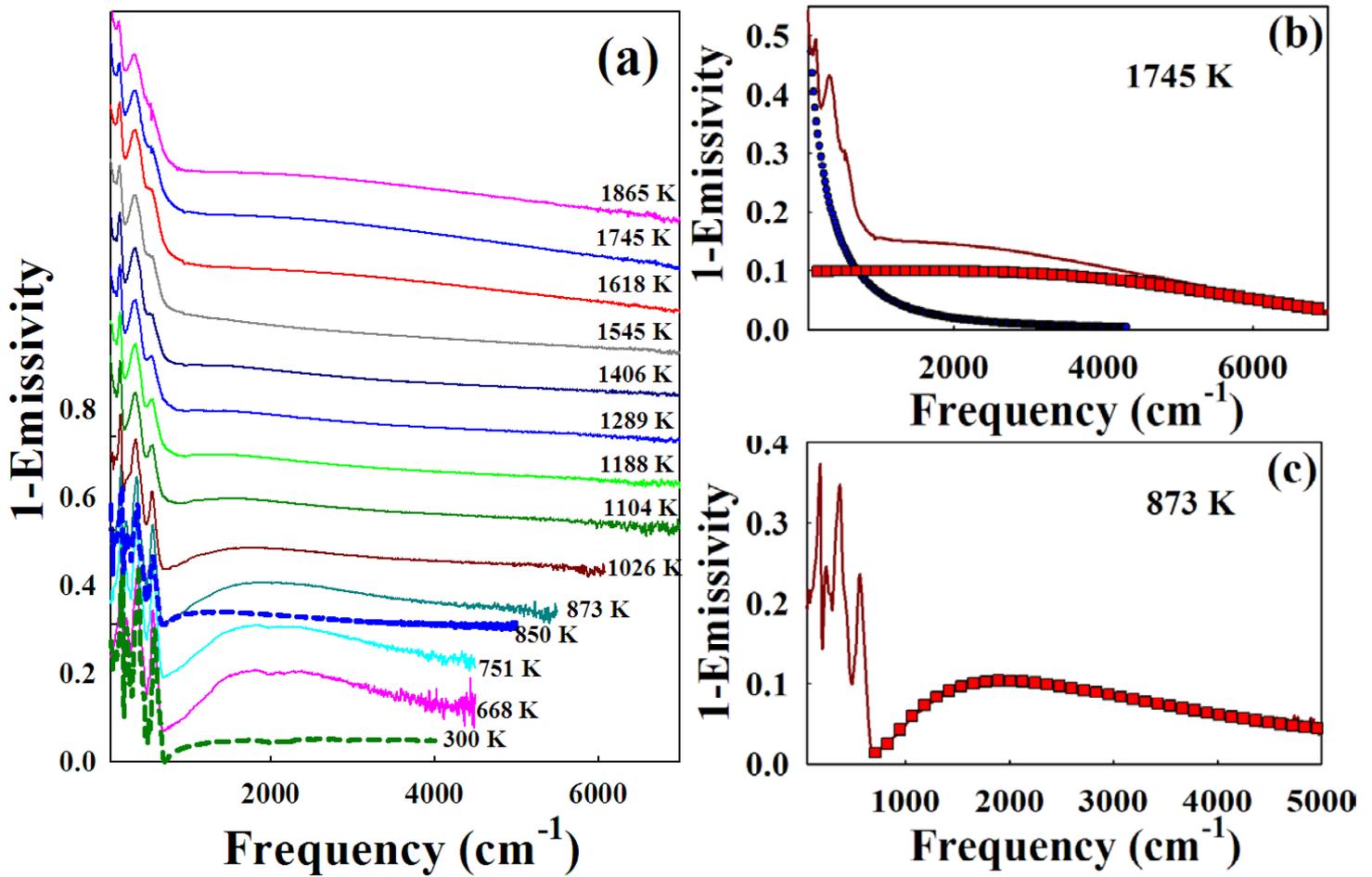

**Figure 4
Massa et al**

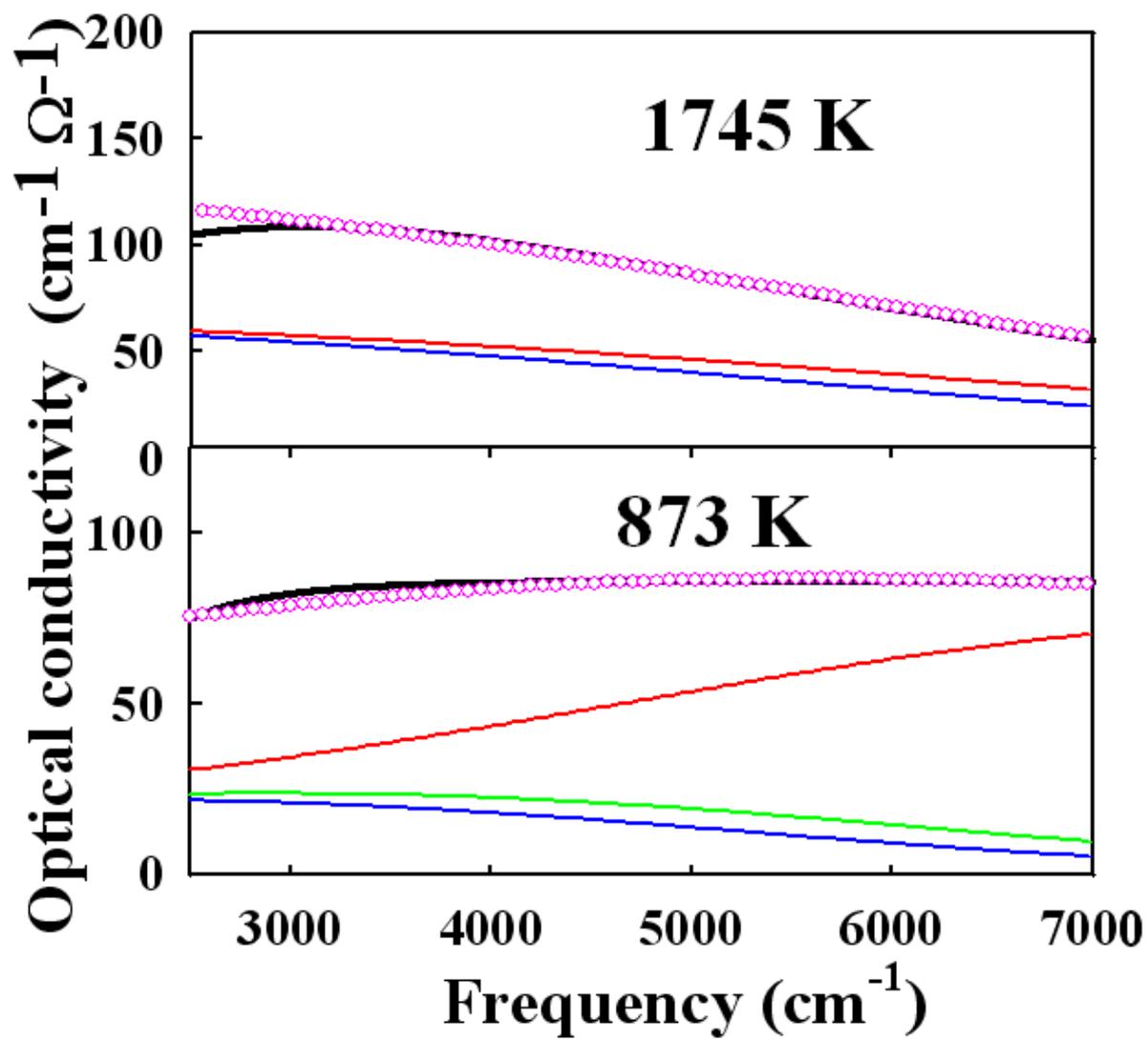

Figure 5
Massa et al